# A simple numerical model of the water bottle flipping experiment

Julie Nassoy[1], Margot Nguyen Huu [1] Léon Rembotte[2,3,4,5], Jean-Baptiste Trebbia[2,3,4], and Pierre Nassoy[2,3,4]

[1] Lycée Montaigne, Bordeaux, France

[2] University of Bordeaux, Laboratoire Photonique Numérique et Nanosciences (LP2N), 33400 Talence, France

[3] Institut d'Optique, LP2N, 33400 Talence, France

[4] CNRS, UMR 5298, 33400 Talence, France

[5] Centre de Recherche Paul Pascal (CRPP), UMR 5031, CNRS and University of Bordeaux, 33600 Pessac, France

E-mail: pierre.nassoy@u-bordeaux.fr

**Abstract**

The water bottle flip experiment is a recreational, non-conventional illustration of the conservation of angular moment. When a bottle partially filled with water is thrown in a rotational motion, water redistributes throughout the bottle, resulting in an increase of moment of inertia and thus to a decrease in angular velocity, which increases the probability of it falling upright on a table as compared with a bottle filled with ice. The investigation of this phenomenom is accessible to undergraduate students and should allow them to gain better understanding of combined translational and rotational motions in classical mechanics. In addition to reporting a series of detailed experiments and analyzing them using standard image analysis, we provide a simple theoretical framework and subsequent numerical implementation based on the decomposition of the water volume into thin slices of a rigid body that are subjected to fictitious forces in the non-inertial frame of the spinning bottle. This model allows us to capture quantitatively the main experimental findings, including the angular velocity slow-down for a given range of water filling fractions. Finally, we discuss additionnal counter-intuitive effects that contribute to bottle stabilization on landing.

**1. Introduction**

The motion of a rigid bodycan always be seen as the combination of a translation of the center of mass $C_M$ that describes a parabolic trajectory and a rotation around $C_M$ [1]. While rotation can often be overlooked to describe the motion of a high-symmetry projectile like a ball, the complex trajectory of an anisotropic body cannot be reduced to the motion of its center of mass. For instance, successful axe or knife throwing not only requires reaching the target but also ensuring that the blade points forward after one or several rotations. The identification of the center of mass of the projectile is thus the prerequisite to decompose the trajectory into translational and rotational components. First, from Newton's second law, the sum of the forces acting on the system at $C_M$ is equal to the time variation in linear momentum. Second, in the frame of the flying body, the sum of the torques of these forces about $C_M$ is equal to the rate of change of the angular momentum about $C_M$. Additional difficulties arise if the coordinates of $C_M$ within the reference frame of the body change during its motion. This situation is remarkably illustrated in the water bottle flip experiment.

This experiment, which started as an award-winning challenge in 2016, consists in making a partially water-filled bottle do a somersault and having it land upright on a table [2]. The motion of the water bottle was analyzed by measuring the angular velocity throughout the trajectory from hand release to landing [3]. A toy model consisting of a tube containing two tennis balls was used to propose that there is a redistribution of mass leading to an increase of moment of inertia, with the two balls moving towards the ends of the tube, and thus to a decrease in the angular velocity according to the law of conservation of angular momentum. By seeing how colored water spreads out within the bottle during the somersault (Fig. 1), a similar effect may qualitatively explain why the measured angular velocity of the water bottle also decreases during movement. While writing the moment of inertia of the water bottle as a function of all geometric parameters (radius, length, mass) and the position of $C_M$ is straightforward [3], the main difficulty actually remains in determining the coordinates of $C_M$ accurately. The approximation that the distribution of mass inside the bottle is uniform is very bold on the basis of the snapshots taken during a typical experiment (Fig. 1).

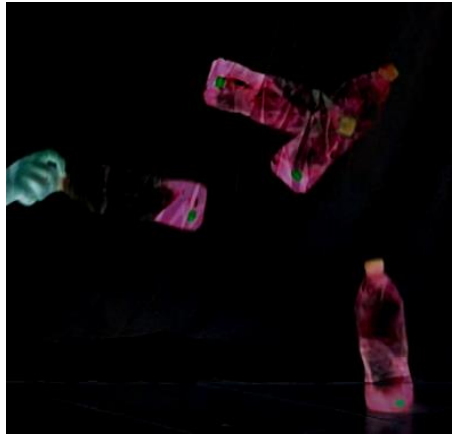

*Figure 1: Composite false-colour image of 4 snapshots taken during water bottle flipping showing that water spreads out all over the bottle during rotation. The green dot at the bottom of the bottle is a marker to facilitate tracking. The bottle is 0.28 m high.*

In this article, we propose a simple numerical model that allows us to compute the coordinates of $C_M$ along the trajectory of the bottle. Instead of truly regarding the fluid nature of the bottle content, we consider the water volume as a rigid body composed of N independent thin solid slices moving along the axis of the bottle. The whole duration of the movement is discretized into P small time intervals. The position of $C_M$, and subsequently the moment of inertia of the bottle, can be computed at each time step. We assess the validity of this model by comparison with dedicated experimental data and with the case of a bottle partially filled with frozen water that can be treated analytically. The limitations of the model will also be discussed.

The concepts of rotating non-inertial frames of reference are classical topics in introductory mechanics courses in college-level education. However, the associated fictitious forces are sometimes difficult to understand by students. Here, they are addressed in the direct context of an entertaining example.

**2. Notations and Methodology**

The laboratory (inertial) reference frame $\mathcal{R}$ is defined by the (O,x,y,z) cartesian coordinate system. (Oxy) is chosen to be the plane of the trajectory. The initial time is taken when the bottle is released from the hand. At t=0, the bottle of length L is represented by the vector $\overrightarrow{OB}$, with the cap being taken as the origin at t=0, and the bottom of the bottle marked by B. The angle between $\overrightarrow{OB}$ and the y-axis is $\theta_0$ at t=0 (see Fig. 2). All notations are summarized in Fig. 2a. In practice, there is a natural tendency from a human experimentalist to release the bottle when $\theta_0 \approx \pi/2$. However, the numerical model provided here will allow us to vary $\theta_0$ and the initial velocity $v_0$.

In the non-inertial reference frame $\mathcal{R}$' associated with the bottle and defined by the polar coordinate system

$(O', \vec{u_r}, \vec{u_\theta}, \vec{u_z})$, the bottle is identified by the vector $\overrightarrow{O'B}$ (with the cap coinciding with the origin) and its angle θ(t) with vertical axis.

Experimentally, image analysis will yield the coordinates of cap C and bottom B in time, from t=0 to t=T (on landing). The angle θ will be directly derived from:

$$\tan \theta = \frac{x_B - x_{O'}}{y_{O'} - y_B} \tag{1}$$

Finally, after numerical derivation of θ(t), the time variation of the angular velocity ω(t) is obtained. Remarkably, the experimental determination of ω(t) does not require the knowledge of the coordinates of $C_M$ in time.

Numerically, we aim instead to derive ω(t) by calculating the moment of inertia J(t) about the moving center of mass of the partially filled bottle and applying the law of conservation of the angular momentum $L(t) = J(t).\omega(t) = J_0.\omega_0$, where $J_0$ and $\omega_0$ are the moment of inertia and angular velocity at t=0. To do so, we need to determine the polar coordinates of $C_M$ in $\mathcal{R}$'. Our approach consists in assuming that the water mass ($M_w$) is decomposed in N slices of mass $M_w/N$ that have no interaction with each other and are free to slide along the main axis of the bottle. This alleviates the need to take rigorous account of the fluid properties of water while allowing to mimic water spreading through the motions of each slice of water taken as an independent solid body. The equations of motion and the implementation of the numerical model will be detailed in section 4.

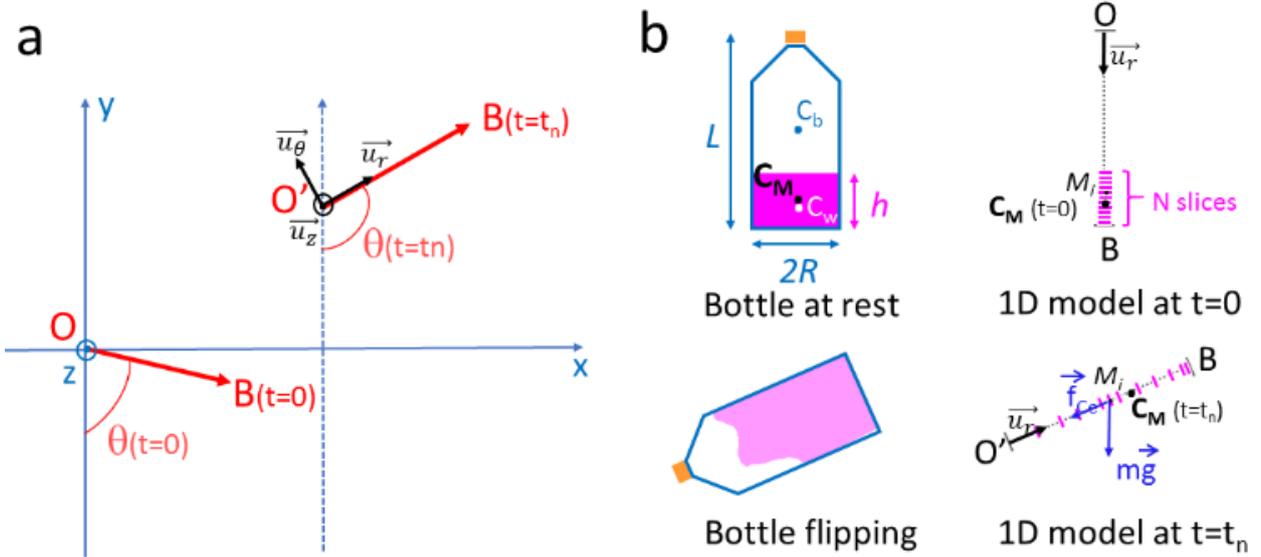

*Figure 2*: *Notations. (a) Representation of the inertial (laboratory) reference frame R (Oxyz) and the non-inertial rotating reference frame associated with the bottle R' $(O', \vec{u_r}, \vec{u_\theta}, \vec{u_z})$; (b) Sketch of the bottle at rest and in motion and the corresponding configurations derived from the numerical 1D model.*

**3. Experimental data and analysis**

We performed a series of experiments to collect data against which our numerical model will be quantitatively compared. The methodological approach is similar to the one developed in [3, 4]. We used a tablet (iPad Pro, Apple) equipped with a camera acquiring 1280x720 pixels images at a rate of 240 frames per second. A plastic bottle was partially filled with colored water, with a varying filling mass fraction $\widetilde{M} = M_w/M_{max}$, where $M_{max}$ is the mass of water corresponding to complete filling (in our case, $M_{max}$=1 kg). Note that, if the bottle is assumed to be a perfect cylinder, we also have $\widetilde{M} = h/L$, where h is the height of water in the bottle at rest.

Each video was analyzed using the *Tracker* software [5]. A sticker was placed at the bottom of the bottle to facilitate tracking. The cap served as a natural second marker at the top of the bottle. A white curtain was placed in the background to improve the efficiency of semi-automatic detection of the markers. Only sequences corresponding to successful flips were selected (Supp videos 1-5). As a control, we used a bottle partially filled with frozen water (Supp videos 6). In this case, the ice pack truly behaves as a rigid body with a fixed center of mass. We thus expect the angular velocity to remain constant throughout the trajectory. The probability of successful trials is significantly reduced with the frozen water bottle and the rare upright landings are often preceded with a rocking motion similar to the one of a roly-poly toy [6]. This effect is however beyond the scope of the present paper.

Qualitatively, compositional pictures of the bottle trajectory were obtained with the *ImageJ* software [7] by using the *Image calculator* macro: the *Min* value of successive images taken in pairs provides a stroboscopic effect and the *Difference* with a background image produces false color images that enhance water mass distribution (Fig.1).

Quantitatively, tracking data $(t, x_{O'}, y_{O'}, x_B, y_B)$ were recorded and exported to the data processing *Origin Pro* software (version 7, RITME, France). First, we validated the methodology with a bottle partially filled with frozen water (Fig. 3a). In this case, we additionally tracked the coordinates of the top (T) surface of the ice pack, $x_T(t), y_T(t)$. By neglecting here the mass of the plastic bottle, the center of mass of the system was taken to be located halfway between top of the ice pack (T) and base of the bottle (B). We immediately observe that, whereas O', B and T exhibit complex trajectories $x_{CM}$ varies linearly with time and $y_{CM}$ follows a parabolic-like time variation, in agreement with the expected kinematics of $C_M$:

$$x_{CM} = v_{0x}t + (L - \frac{h}{2}) \sin \theta_0 \tag{2a}$$

$$y_{CM} = -\frac{1}{2}gt^2 + v_{0y}t - (L - \frac{h}{2}) \cos \theta_0 \tag{2b}$$

The fit of the data with Eq. (2a) and (2b) is excellent and yields for instance $v_0 = \sqrt{v_{0x}^2 + v_{0y}^2} \approx 2$ m.s$^{-1}$, which we will use as an order of magnitude for $v_0$ in the numerical model. Fig. 3a also displays representative plots of $x_{O'}(t), y_{O'}(t), x_B(t), y_B(t)$ for three filling fractions of liquid water. These curves are quite diverse and difficult to interpret as they are. Fig. 3b shows a more visual representation of the trajectory based on a vector plot, where vector tail is the cap and vector direction is obtained from Eq. (1).

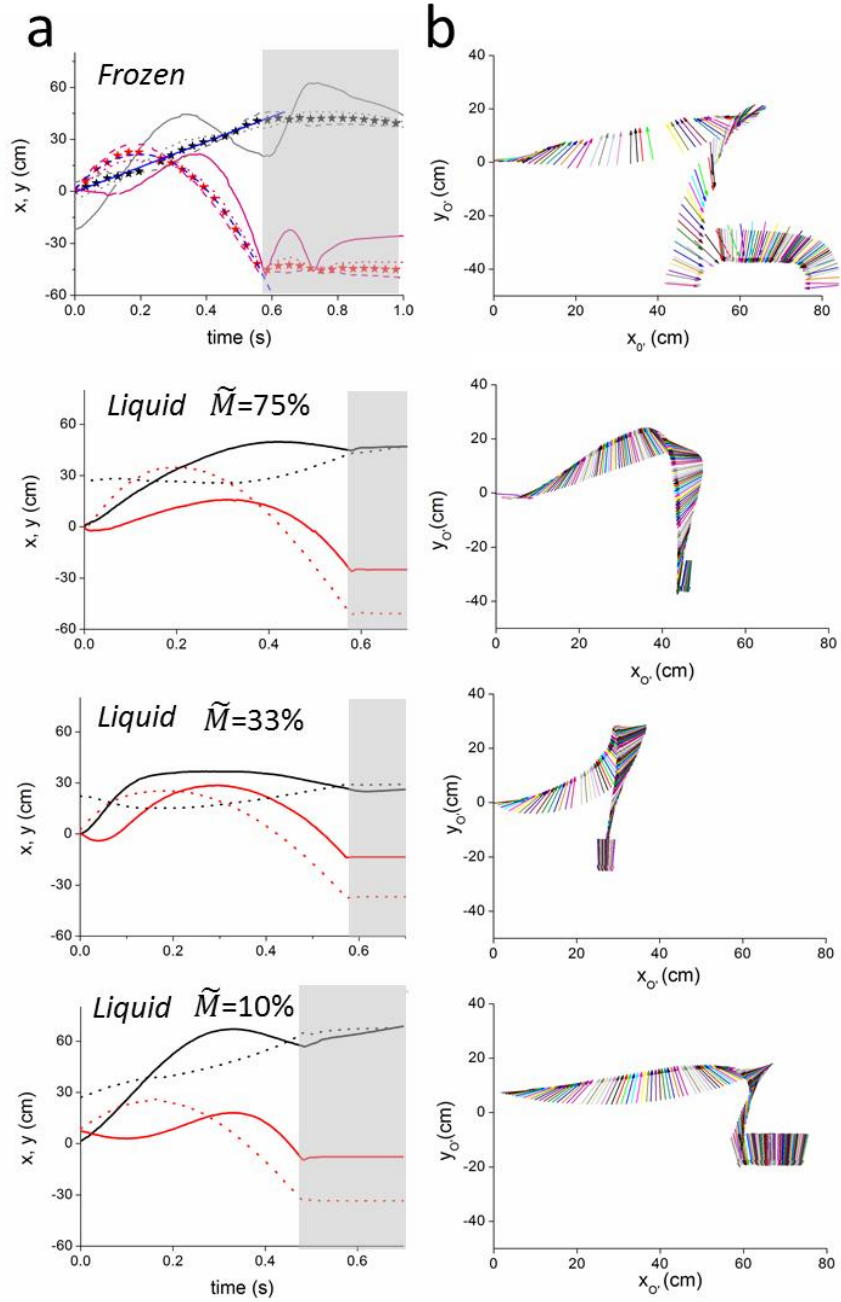

***Figure 3***: *Representative experimental data: **(a)** Coordinates x (black lines) and y (red lines) of the cap (plain) and the bottom of the bottle (dotted) as a funtion of time for the control frozen water bottle (top) and for 3 filling mass fractions of liquid water. For the frozen water case, the top of the ice pack surface is represented in dashed line and the calculated center of mass as stars. Linear and parabolic fits using Eq. (2a) and (2b) are shown in blue. The transparent grey portion of the graphs marks the moment when the bottle touches the ground. In all liquid cases, the bottle lands upright. In the case of the frozen bottle, the oscillations that follow correspond to a roly-poly effect: the bottle lands flat and bounces upright; **(b)** Vector plot representation, with O'coinciding with vector tail.*

This vector plot implicitly required to calculate θ(t) using Eq.(1). Direct examination of the videos reveals that the spin of the bottle exceeds half a revolution, meaning that θ(t) becomes larger than π. Since θ(t) is derived from its tangent, we expect discontinuities the vicinity of π/2. The singularities were corrected manually by adding π when necessary. Fig. 4a-b show representative plots of θ(t), which is measured, and ω(t), which is obtained by calculating numerically the time derivative of θ(t) after interpolation (in case of missing points due to tracking errors) and smoothening (by averaging over 5 points) for the frozen water case and the liquid water case with $\widetilde{M} = 0.33$. Since all cases were successful tosses, the terminal values of θ(t) are close to 0 (mod2π). More important is the evolution of the the angular velocity during the motion. It remains constant, around 20 rad.s$^{-1}$ for the frozen water experiment, while ω(t) exhibits two phases for the liquid water bottle before landing associated with a sudden drop of ω(t) to zero: it first decays quickly and then remains constant at a lower value. The observed evolution of ω(t) is consistent with the previous report in [3], even though only normalized data were given. Here, we purposely reported actual measurements. The first point of validation of the following numerical model will be to recover the order of magnitude of the angular velocity. Then, we explored whether this behavior was found for all filling fractions. Fig. 4b shows the normalized the angular velocity to its initial value $ω_0$, $\widetilde{ω} = ω/ω_0$, as a function of the the time to the landing time T, $\widetilde{t} = t/T$. Fig 4d summarizes the results for different $\widetilde{M}$ values. We observe that $\widetilde{ω}$ is minimal for filling fractions between 0.25 and 0.50. Nontheless, there is already a significant reduction in $\widetilde{ω}$ for $\widetilde{M} = 0.75$. Contrastingly, the variation of $\widetilde{ω}$ for $\widetilde{M} = 0.10$ is very close to the one of the frozen water bottle, indicating that there is no significant redistribution of the water mass inside the bottle during the flip, which is indeed seen in the videos. In other words, if the center of mass is initially too low, the centrifugal forces tend to keep water at the bottom of the bottle. For filling fractions close to 1, the center of mass is already quite high (roughly at the center of the bottle), and water redistribution is not accompanied with a large displacement of the center of mass. We thus expect an optimal filling, leading to a maximal increase in the moment of inertia, and thus to the maximal slowdown in angular velocity.

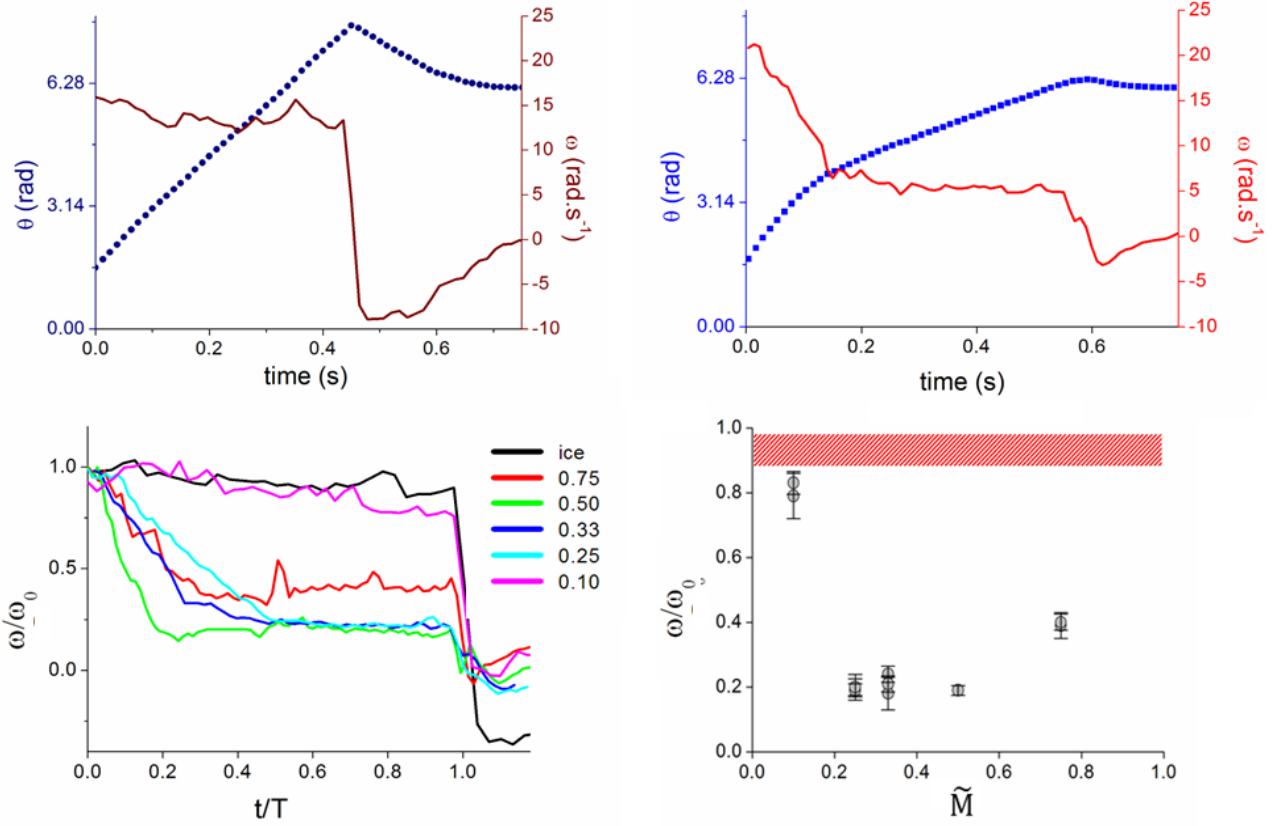

*Figure 4*: Representative variation of the angle and angular velocity of the bottle during motion for **(a)** frozen water, and **(b)** liquid water, with a filling fraction of 0.33; **(c)** Normalized angular velocity as a function of normaized time for frozen and liquid water at different filling fractions: $\widetilde{M}$ =0.75; 0.50; 0.33; 0.25 and 0.10; **(d)** Average $\widetilde{\omega} = \omega/\omega_0$ versus $\widetilde{M}$. The red shaded area corresponds to the frozen water case.

### 4. Theoretical framework

The water content of the bottle is modeled as composed of N identical solid slices that may slide along the main axis of the bottle. We further simplify the system by considering it as pseudo 1D given the cylindrical symmetry.

As shown in Fig. 2b, at t=0, all identical slices of mass $m = M_w/N$ are stacked at the bottom of the bottle. Let us note $M_i$ the center of mass of slice i. At t=0, for $1 \leq i \leq N$, the position of $M_i$ is defined by $\vec{r_i} = \overrightarrow{O'M_i}$ and given by:

$$\vec{r_i}(t=0) = \left[L - (2i-1)\cdot\frac{h}{2N}\right]\vec{u_r} \tag{3a}$$

The center of mass $C_W$ of the water content is located half-way between bottom and top of the water surface and

$\overrightarrow{r_{Cw}} = \overrightarrow{O'C_w}$ is given by:

$$\overrightarrow{r_{Cw}}(t=0) = \left[L - \frac{h}{2}\right]\overrightarrow{u_r} \tag{3b}$$

For high filling fractions, the mass of the bottle (typically 20-30 grams) can be neglected to determine the position of the center of mass of the whole "Bottle+Water" system. However, this approximation does not hold for low filling fractions. We have thus to take into account the mass of the bottle, $M_b$. Its center of mass $C_b$ is an invariant defined by $\overrightarrow{r_{Cb}} = \overrightarrow{O'C_b}$:

$$\overrightarrow{r_{Cb}}(t=0) = \overrightarrow{r_{Cb}}(t>0) = \frac{L}{2}\overrightarrow{u_r} \tag{3c}$$

The total center of mass of the system is then identified by $\overrightarrow{r_{CM}} = \overrightarrow{O'C_M}$ and obtained by a weighted average of both sub-systems:

$$\overrightarrow{r_{CM}}(t=0) = \frac{M_b\overrightarrow{r_{Cb}}(t=0) + M_w\overrightarrow{r_{Cw}}(t=0)}{M_b + M_w} \tag{3d}$$

Projection on $\overrightarrow{u_r}$ reads:

$$r_{CM}(t=0) = \left(1 - \frac{\epsilon}{2}\right)L - (1-\epsilon)\frac{h}{2} \tag{3e}$$

with $\epsilon = \frac{M_b}{M_b + M_w}$ \hfill (4a)

By assuming that the bottle is a hollow cylinder of radius R, $\epsilon$ rewrites:

$$\epsilon = \frac{H_0}{H_0 + h} \tag{4b}$$

where $H_0$ is defined by:

$$H_0 = \frac{M_b}{\rho_w \pi R^2} \tag{5}$$

With $\rho_w$ the density of water, and R the radius of the bottle.

Note that Eqs (3d) and (3e) remain valid at any time t.

At time t, the slices are spread out over the entire length of the bottle (Fig. 2b). In the non-inertial reference frame $\mathcal{R}'$, we apply the modified Newton's second law of motion to the slices i defined by their centers of mass $M_i$ by introducing the fictitious forces, namely the centrifugal (or inertial) force, $\overrightarrow{f_{Ce}}$, the Coriolis force, $\overrightarrow{f_{Co}}$, and the Euler (or azimuthal) force $\overrightarrow{f_{Eu}}$ [8]:

$$m\vec{a}_{M_i/\mathcal{R}'} = m\vec{g} + \overrightarrow{f_{Ce}} + \overrightarrow{f_{Co}} + \overrightarrow{f_{Eu}} + \overrightarrow{f_D} \qquad (6),$$

With $m\vec{g}$ the weight of one slice, and $\overrightarrow{f_D}$ a phenomenological drag force proportional to the velocity of the water slices and their mass m, and characterized by a drag coefficient α that is determined empirically.

We need to adapt the general expression of the fictitious forces, which are usually given by taking the axis (O'z) as the axis of rotation. Here, the rotation takes place about the axis ($C_M$, z) which slides along the bottle axis as $C_M$. We then have:

$$\overrightarrow{f_{Ce}} = -m\vec{\omega} \times (\vec{\omega} \times \overrightarrow{C_M M_i}) \qquad (7a)$$

$$\overrightarrow{f_{Co}} = -2m\vec{\omega} \times \left(\frac{d\overrightarrow{C_M M_i}}{dt}\right)_{\mathcal{R}'} \qquad (7b)$$

$$\overrightarrow{f_{Eu}} = -m\left(\frac{d\vec{\omega}}{dt}\right)_{\mathcal{R}'} \times \overrightarrow{C_M M_i} \qquad (7c),$$

with $\overrightarrow{C_M M_i} = \vec{r_i} - \vec{r_{CM}}$

Note that the Euler force is often omitted. In the particular context of the water bottle fliping experiment, the variation of $\omega(t)$ in time is key for explaining the underlying physics. However since there is only a variation in magnitude and not in direction of $\omega(t)$, $\overrightarrow{f_{Eu}}$ is tangential to the motion and therefore perpendicular to the main axis $\vec{u_r}$. Similarly, $\overrightarrow{f_{Co}}$ is also oriented along $\vec{u_\theta}$. To determine the distribution of mass inside the bottle, we only need to examine the motion of the $N_s$ slices along the bottle axis. Projection of (6) on $\vec{u_r}$ gives:

$$m\ddot{r}_i = mg.\cos\theta + m\omega^2(r_i - r_{CM}) - \alpha(\dot{r}_i - \dot{r}_{CM}) \qquad (8)$$

Eq. (8) defines the kinematics of the bottle motion.

## 5. Numerical model

Classical mechanics teach us that solutions of Eq.(8) take the form:

$$r_i(t) - r_{CM}(t) = (Ae^{\omega t} + Be^{-\omega t})e^{-(\alpha/2m)t} - \frac{g}{\omega^2}\cos\theta \qquad (9a),$$

with A and B determined by the boundary conditions and by assuming $\frac{\alpha}{2m} = \frac{\alpha N}{2M_W} \ll \omega$ and ω=constant. The latter assumption obviously seems invalid over the duration of the motion since we precisely aim to show that $\omega$ varies. However, by discretizing the duration T of the bottle flight into $N_t$ short time intervals $\Delta t = t_n - t_{n-1}$ (for n=1 to $P = T/\Delta t$), this assumption will hold over one time step $\Delta t$ if $\omega$ has slow variations, which is indeed found experimentally for small enough $\Delta t$. As a corollary, $\theta(t)$ and $r_{CM}(t)$ are also expected to vary slowly over one time step and will thus be taken as constants in a time interval $\Delta t$ as a first approximation.

The boundary conditions are determined from the values of $r_i(t_{n-1})$, $r_{CM}(t_{n-1})$, $\dot{r}_i(t_{n-1})$, $\omega_{n-1} = \omega(t_{n-1})$ and $\theta_{n-1} = \theta(t_{n-1}) = \theta_{n-2} + \Delta t \cdot \omega_{n-1}$ calculated at the beginning of each time step $[t_{n-1}; t_n]$ and derived by recurrence from initial time $t_0$. For the initial positions of the water slices i and the center of mass, we use Eqs.(3a) and (3e). The slices i are at rest at $t_0$, setting $\dot{r}_i(t_0) = 0$. The initial angular velocity is $\omega_0 = v_0/r_{CM}(t=0)$ and $\theta_0$ is chosen equal to $\pi/2$ unless otherwise stated.

Eq. (7a) then becomes for $t \in [t_{n-1}; t_n]$:

$$r_i(t) = \left(Ae^{\omega(t-t_{n-1})} + Be^{-\omega(t-t_{n-1})}\right)e^{-\left(\frac{\alpha}{2m}\right)(t-t_{n-1})} - \frac{g}{\omega^2}\cos\theta(t_{n-1}) + r_{CM}(t_{n-1}) \tag{9b}$$

with A and B constants.

Finally, we obtain the locations and velocities of the N slices (i=1 to N) at time $t_n$ (n=1 to P):

$$r_i(t_n) = \left[r_i(t_{n-1}) - r_{CM}(t_{n-1}) + \frac{g}{\omega^2}\cos\theta_{n-1}\right].e^{-\left(\frac{\alpha}{2m}\right)(t-t_{n-1})}\cosh\omega_{n-1}(t_n - t_{n-1}) + \left[\frac{\dot{r}_i(t_{n-1})}{\omega} + \frac{\frac{\alpha}{2m}}{\omega}\left(r_i(t_{n-1}) - r_{CM}(t_{n-1}) + \frac{g}{\omega^2}\cos\theta_{n-1}\right)\right].e^{-\left(\frac{\alpha}{2m}\right)(t-t_{n-1})}\sinh\omega_{n-1}(t_n - t_{n-1}) + r_{CM}(t_{n-1}) - \frac{g}{\omega^2}\cos\theta_{n-1} \tag{9c}$$

and

$$\dot{r}_i(t_n) = \left[\frac{\omega^2 - (\alpha/2m)^2}{\omega}\left(r_i(t_{n-1}) - r_{CM}(t_{n-1}) + \frac{g}{\omega^2}\cos\theta_{n-1}\right) + \frac{\frac{\alpha}{2m}}{\omega}\dot{r}_i(t_{n-1})\right].e^{-\left(\frac{\alpha}{2m}\right)(t-t_{n-1})}\sinh\omega_{n-1}(t_n - t_{n-1}) - \dot{r}_i(t_{n-1}).e^{-\left(\frac{\alpha}{2m}\right)(t-t_{n-1})}.\cosh\omega_{n-1}(t_n - t_{n-1}) \tag{9d}$$

Eq. (9c) and (9d) govern the motions of the rigid slices inside the non-inertial reference frame of the bottle. To better grasp their physical meaning, the motion is controlled by the balance between the centrifugal force and gravity. If the initial thrust is too small, the bottle will just fall. Eq. (8) gives the order of magnitude of the angular velocity: $\omega \sim \sqrt{g/\ell} \sim 10 \ rad.s^{-1}$ where $\ell$ is the characteristic length of the system, of the order of $L/2 \simeq 0.1$m. Note however that the more distant the slice of interest and the centre of mass, the larger the magnitude of the centrifugal force. In other words, slices that are next to the bottom or the top of the bottle will be pushed further against the bottom and the top of the bottle, respectively. Eqs. (9c) and (9d) do not include this boundary condition: water slices cannot go below 0 and above L, or more precisely, their center of mass are restricted to the $[h/2N, L - h/2N]$ interval. Moreover, the present model does not account for interaction beween slides, meaning that water slices can overlap. However, to avoid accumulation of all water slices either at the bottom or the top of the bottle and to account for water sloshing, we intoduce a rebound of the water slices when they hit the top and bottom of the bottle. Numerically, we introduce the following conditional statement:

[C.S] if $r_i(t_n) \leq h/2N$ (slice i hits the cap) or $r_i(t_n) \geq L - h/2N$ (slice i hits the bottom) then $r_i(t_n) = r_i(t_{n-1})$ *(slice i does not move)* and $\dot{r}_i(t_n) = -\beta \dot{r}_i(t_n)$ *(slice i bounces back in an elastic ($\beta = -1$) or inelastic ($-1 < \beta \leq 0$) collision with the bottle bottom or top).*

At each time step, after computing $r_i(t_n)$, the coordinate of the center of mass of the "Bottle+Water" system is readily derived using Eq. (3c) and the definition of $C_M$:

$$r_{CM}(t_n) = (1-\varepsilon) \sum_{i=1}^{N} \frac{1}{N} r_i(t_n) + \varepsilon \frac{L}{2} \tag{10}$$

Knowing the position of $C_M$, the next step is to compute the moment of inertia J of the water bottle about $C_M$. The bottle itself, which is treated as a hollow cylinder of length L, radius R and mass $M_b$ has a moment of inertia about its center of mass $C_b$:

$$J_{b/C_b} = M_b \left( \frac{R^2}{2} + \frac{L^2}{12} \right) \tag{11a}$$

However, the axis of rotation of the system passes through $C_M$. Using the parallel-axis theorem [8], the moment of inertia of the bottle about $C_M$ becomes:

$$J_{b/C_M}(t_n) = M_b \left[ \frac{R^2}{2} + \frac{L^2}{12} + \left( \frac{L}{2} - r_{CM}(t_n) \right)^2 \right] \tag{11b}$$

Regarding the water mass content, at $t = t_n$, the moment of inertia about $C_M$ is:

$$J_{w/C_M}(t_n) = \sum_{i=1}^{N} \frac{M_w}{N} \left( r_i(t_n) - r_{CM}(t_n) \right)^2 \tag{11c}$$

with $r_i(t_n)$ and $r_{CM}(t_n)$ given by Eq. (9c) and (10) respectively.

Finally, the total moment of inertia during water bottle flipping at time $t_n$ is given by:

$$J_n = J(t_n) = J_{w/C_M}(t_n) + J_{b/C_M}(t_n) \tag{11d}$$

The moment of inertia around $C_M$ is therefore calculated by recurrence as for $r_{CM}(t_n)$.

The initial value $J_0$, which is calculated by using the discrete definition of the moment of inertia and the expression of $r_{CM}(t=0)$, can be brought to the form:

$$J_0 = M_b L^2 \left[ \frac{1}{12} + \frac{R^2}{2L^2} + \left( \frac{\varepsilon}{2} + (1-\varepsilon)\frac{\widetilde{M}}{2} \right)^2 \right] + M_w L^2 \left[ \widetilde{M}^2 \left( \frac{1}{3} - \varepsilon + \varepsilon^2 \right) - \widetilde{M} \varepsilon \left( 1 - \frac{\varepsilon}{2} \right) + \frac{\varepsilon^2}{4} \right] \tag{11e}$$

From angular moment conservation, we then derive the angular velocity during bottle flip:

$$\omega_n = \frac{J_0 \cdot \omega_0}{J_n} \tag{12}$$

## 5. Simulated water bottle flipping experiments

We have developed a MATLAB code and have included both the MATLAB script and its corresponding Python program (shared for open-source purposes) as Supplementary Text 1. The primary goal is to plot the angular velocity just before landing of the bottle as a function of the filling fraction and to compare these computed values with the experimental data from Figure 4. To do so, we first set the values of known parameters: bottle radius R=0.04 m, height of the bottle L=0.28 m, density of water $\rho_w$=1000 kg.m$^{-3}$, bottle mass $M_b$=0.028 kg, Measured parameters $\theta_0 = \pi/2$ and $\omega_0 = 20\ rad.s^{-1}$ were also set. The two phenomenological unknown parameters in the mathematical model are α and β.

For given drag coefficient α and filling fraction $\widetilde{M}$, β sets the boundary conditions for the velocity of water slices. We first investigate the impact of β on the trajectory of the slices and the center of mass of the bottle (Fig. 5a) and, subsequently on $\theta(t)$ (Fig. 5b) and $\omega(t)$ (Fig. 5c). β=0 corresponds to the case where all slices that reach the ends of the bottle cannot be pushed further and remain stuck at this position. Only when the projection of the weight of a slice along $\overrightarrow{u_r}$ exceeds the centrigugal force can the slices be redistributed. (Fig. 5a, top). In the opposite case of an elastic collision with a rebound of the slices without loss of speed (β=−1, Fig. 5a, medium), we obtain multiple reflections making water slices bounce at the top and bottom of the bottle. This leads to a faster increase of the angle θ in the late rotating stages (Fig. 5b), which is not observed experimentally. Finally, an intermediate value, β=−0.5, seems to avoid accumulation of slides at the ends of the bottle while not amplifying the bounce effect (Fig. 5a, bottom). The difference seen in $\theta(t)$ for β=0 and -0.5 seems to be minor. However, $\omega(t)$ only remains low and constant, without late increase, for β=−0.5. Note the bumpy and irregular variation of $\omega(t)$ when rebounds are elastic. Guided by the observation of these singularities in $\omega(t)$, we more generally investigated the influence of the number of slices $N_s$ on the outcome. Previous results were obtained with N=10. Here, the shapes of the $\omega(t)$ curves only display significant differences for N=2 and N=5. Nonetheless, slight differences in the minimum value of $\omega$ can be detected for N<100.

Since the viscous force $f_D$ leads to a damping with a time constant in the order of $\frac{2M_W}{\alpha N}$, indicating that the drag coefficient is proportional to the "thickness" of a slice, we need to ensure that slices have all the same thickness whatever the filling fraction. In other words, $\frac{N}{M_W}$ was taken constant for all $\widetilde{M}$ values. To avoid loss of precision, we selected $\frac{M_W}{N} = m$=10$^{-3}$ kg. As shown in Supp. Fig. 1, taking $N = 10^3 \widetilde{M}$ is not the main limitation in terms of computation time. The most critical parameter is the number of time intervals P. Computing time $t_c$ roughly increases as N$^2$. However, even for $N \times P = 1000 \times 1000$, $t_c$ does not exceed ~40 s. As a consequence, we chose to run all our calculations with P=1000 time steps.

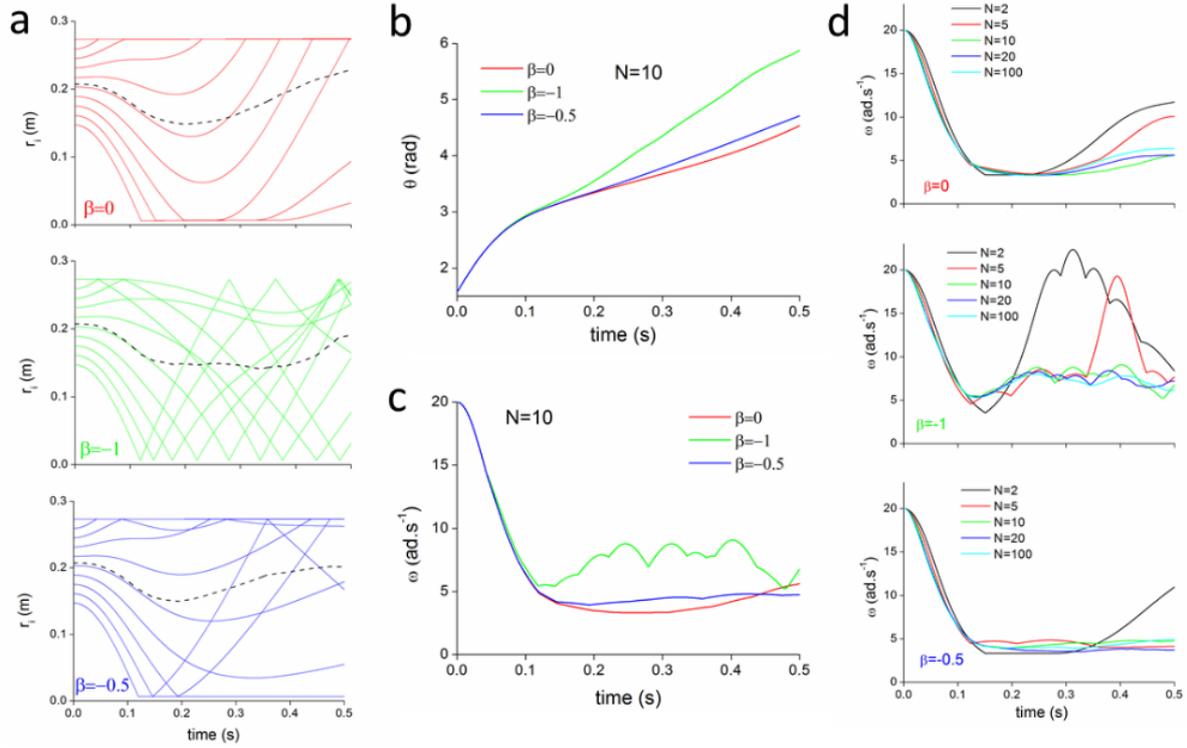

***Figure 5****: Influence of the parameter β describing the boundary conditions for the velocity on **(a)** the trajectories of water slices The dashed black line corresponds to the trajectory of the centre of mass; **(b)** the angle $\theta(t)$ ; **(c)** the angular velocity $\omega(t)$. N=10 slices, $\widetilde{M} = 0.5$, and α =1. The impact of tne number of slices N on the curves $\omega(t)$ is shown in **(d)**.*

Taking β=−0.5 and $m=10^{-3}$ kg , we evaluate the influence of α on the trajectories of the 1$^{st}$, N/2$^{th}$ and N$^{th}$ slices (Fig. 6a), and further the variations of $\theta(t)$ (Fig. 6b) and $\omega(t)$ (Fig. 6c). The outcomes are found to be highly sensitive to α. For α ≥ 0.0125, friction is so large that redistribution of water through the whole bottle is impeded, leading to a non-monotonic variation of the bottle angle and some oscillations in $\omega(t)$. Conversely, if friction is abolished (α = 0), the early stages of flipping in $\theta(t)$ and $\omega(t)$ are well recapitulated, but exhibit later increase which is not observed experimentally. Finally, α values between 0.0075 and 0.01 yield curve shapes similar to the experimental ones (see Fig. 4c) with a flat plateau after a sudden drop for $\omega(t)$.

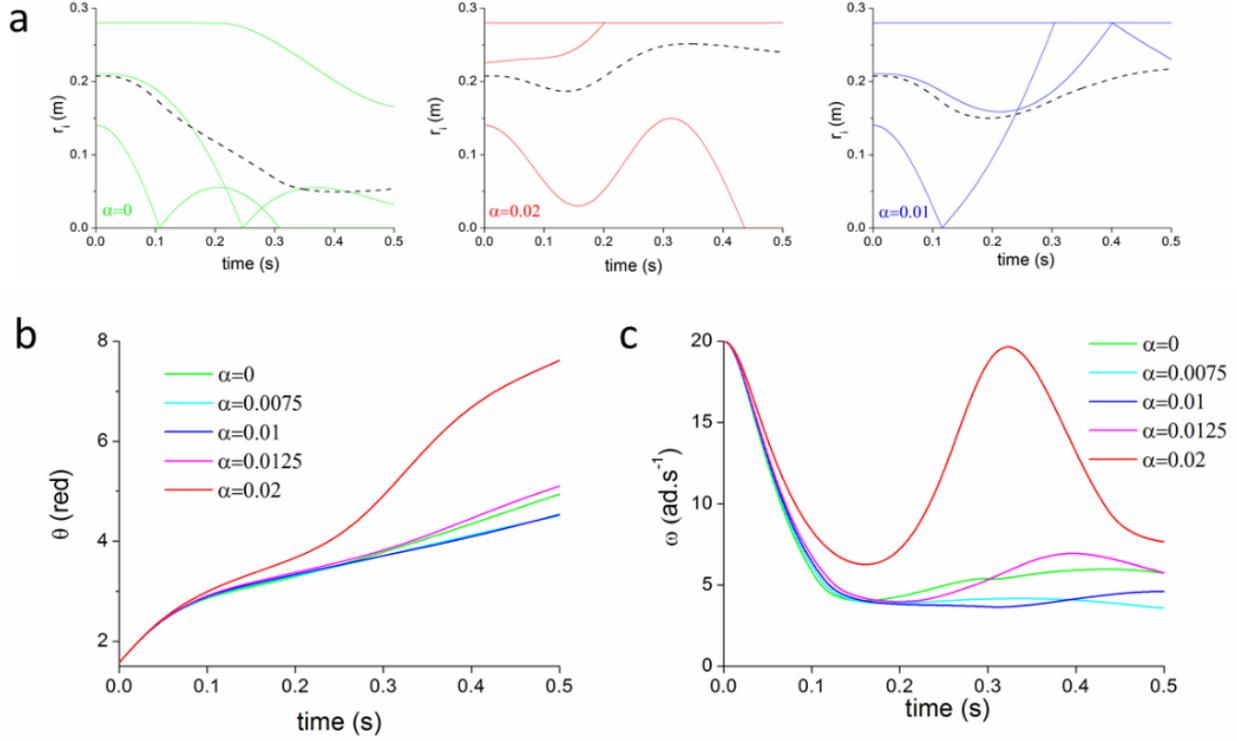

*Figure 6*: Influence of the drag parameter $\alpha$ on (**a**) the trajectories of water slices ($1^{st}$, $250^{th}$ and $500^{th}$). The dashed black line corresponds to the trajectory of the centre of mass; (**b**) the angle $\theta(t)$ ; (**c**) the angular velocity $\omega(t)$. N=500 slices, $\widetilde{M} = 0.5$, and $\beta$ =-0.5.

Finally, we have varied the water filling fraction using the values determined empirically for α=0.01, m=$10^{-3}$ kg and β=-0.5. The initial values of the angle and angular velocity were fixed : $\theta_0 = \pi/2$ and $\omega_0 = 20 \ rad.s^{-1}$. The flipping duration from release to landing was T=0.5s. Figure 7a shows the dimentsionless angular velocity $\omega/\omega_0$ as a function of dimensionless time $t/T$ for $0.05 \leq \widetilde{M} \leq 1$. If we denote $\omega_f$ the plateau value of $\omega(t)$ obtained by calculating the average of $\omega$ over at least 20 time points around the minimum in the pseudo-plateau window, we observe that, in most cases, $\omega_f/\omega_0$ falls in the 0.2-0.5 range for intermediate filling fractions, as observed experimentally (Fig. 7b).

However, for low and high $\widetilde{M}$ values, simulations and experiments show significant differences. First, even though all our tosses were failures for fully filled bottles ($\widetilde{M}$=1), we may anticipate that complete filling prevents water from sloshing around, the moment of inertia from increasing and thus the angular velocity from decreasing. It is expected to be similar to a bottle filled with fozen ice. Yet, we do obtain a decrease in $\omega$ (black curve in Fig. 7a), which may indicate a limit of the validity of the model since all slices are considered as independent and allowed to overlap, in particular with the possibility of accumulating a few slices at one end of the bottle. Second, the measured angular velocity for $\widetilde{M}$=0.1, which displays no decrease during flipping is similar to the simulated one for $\tilde{M}=0.05$, while a net decrease in $\omega(t)$ is obtained for $\tilde{M}=0.1$ from the computational model. The reason why angular velocity is not affected by spin at ultra-low filling fractions is that the mass of the bottle is no longer negligible compared to the water mass and thus the center of mass of the "Bottle+Water" system is less sensitive to water motions, leading to weak variations in moment of inertia. The fact that this change in behavior occurs experimentally for $0.1 \leq \tilde{M} \leq 0.2$ and numerically for $0.05 \leq \tilde{M} \leq 0.1$ is therefore not a critical discrepancy. Altogether, when the plateau value $\omega_f$ of the angular velocity is plotted as a fiunction of $\tilde{M}$, the agreement between experimental (Fig. 4d) and numerical (Fig. 7b) data is excellent, allowing us to define a range of optimum filling between 0.2 and 0.5 to ensure that the rotation speed of the bottle on landing is as low as possible. Note that this range of $\tilde{M}$ also corresponds to the cases for which the highest number of successful tosses was obtained, even though comparison with success probability cannot be rigorously pursued because tosses were done by the authors of the present paper. More controlled launches performed by robots for instance would be preferred.

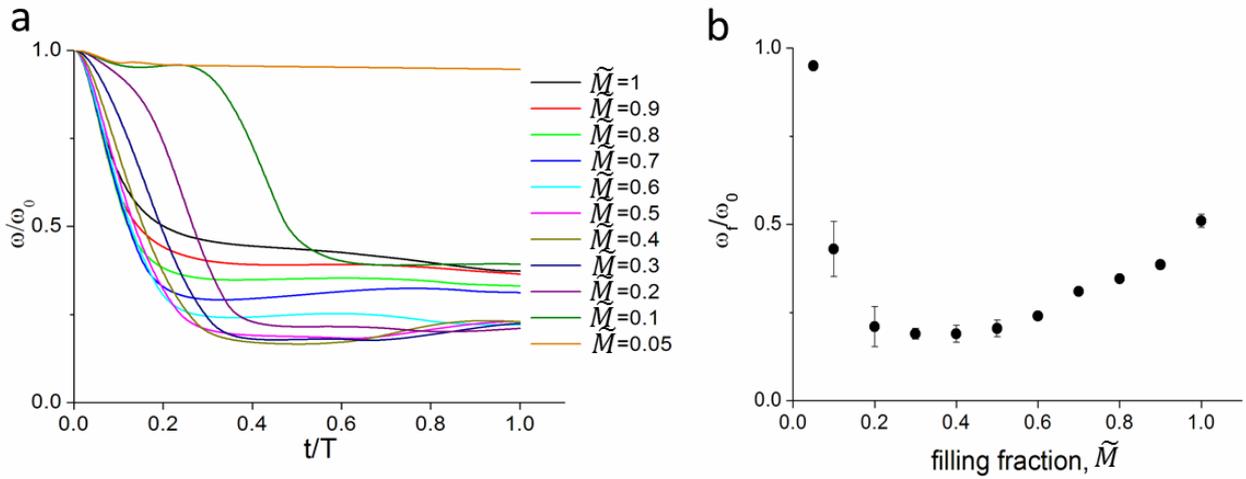

*Figure 7: Influence of the filling fraction on the angular velocity. (a) Variation of the dimensionless angular velocity $\omega/\omega_0$ as a function of dimensionless time t/T for filling fractions varying from 0.05 to 1; (b) Dimensionless $\omega_f/\omega_0$ taken on the plateau as a function of filling fraction.*

Since our numerical model is widely validated over a large range of filling fractions, we further used it to explore the impact of parameters that are not easy to vary by human experimentalists. We observe that tosses are generally performed with an initial velocity close to $\omega_0 \approx 20$ rad.s$^{-1}$ and a release angle $\theta_0 \approx \pi/2$ We thus run simulations by varying these parameters only, for $\tilde{M} = 0.5$ (Fig. 8a and b). We find that tosses at $\omega_0 > 20$ rad.s$^{-1}$ do not affect the final normalized angular velocity, but only the time to reach it. Conversely, tosses at $\omega_0 < 20$ rad.s$^{-1}$ exhibit non monotonic features, suggesting that centrifugal forces become insufficient to redistribute water and clearly alter the moment of inertia for the whole duration of the movement. The impact of the initial release angle is very weak on the overall shape of $\omega(t)$. The decrease time is all the more longer than the bottle is realeased close to the vertical ($\theta_0 = 0$) because water redistribution, which is governed by the competition between centrifugal force and weight, is delayed. Finally, if the aspect ratio AR of the bottle is changed while keeping its volume (or total mass) constant (Fig. 8c), we observe that $\omega(t)$ curves are almost superimposed. However, we may expect that the stability upon landing may differ. We discuss this effect in the next section.

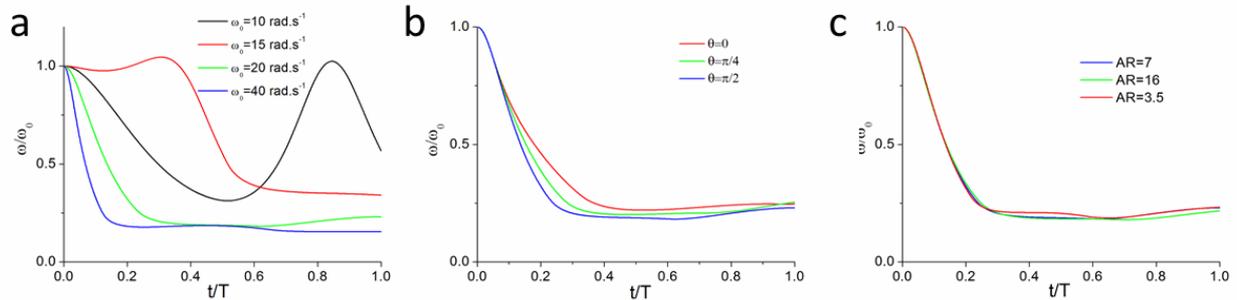

*Figure 8: Normalized angular velocity versus normalized time when (**a**) varying $\omega_0$, (**b**) varying $\theta_0$, and (**c**) aspect ratio of the bottle AR while keeping the same volume.*

## 6. Discussion and additional arguments for successful water bottle flips

The pseudo 1D numerical model proposed here allows to capture the main features of the rotation of water-filled bottles landing upright. The model, which consists in decomposing the water content in thin rigid body slices and describing their motion in the non-inertial frame of the bottle, allows to account for the decrease in the angular velocity as a consequence of a redribution of water within the bottle and thus an increase of the moment of inertia of the whole "Bottle+Water" system. We also recapitulate quantitatively the variations of $\omega(t)$ as compared to the experimental data, by adjusting two parameters, namely the drag coefficient for one slice and the boundary condition for the velocity of the slices at the end of the bottles. The model was proved to be more accurate for partially filled bottles of water. Slight deviation is found for ultra-low filling fraction for which the mass of the empty bottle is no longer negligible. The model breaks down for complete filling, which does not exhibit in principle any variation of moment of inertia by contrast whith what is found in the model where overlapping slices are not ruled out.

The reduction of $\omega$ during bottle rotation was the focus of the present work allowing to investigate an unusual case of mechanics in non-inertial frames. Indeed, if the angular velocity of the spinning bottle is up to 5 times lower than that of a bottle partially filled with a rigid body with a fixed moment of inertia, the probability of maintaining an orientation close to vertical ($\theta \approx 6\pi$), if roughly reached at the apex of the trajectory, will be higher in the last moments preceding landing and almost corresponding to free fall. The angle reached on landing does not need to be exactly $2\pi$. As reported for instance in [9], it is well known that there exists a maximal angle over which a tilted bottle will topple. By assuming that water remains static and its surface parallel to the tilted base of the bottle, a bottle partially filled with fluid will return to its upright position as long as the gravity line does not exceed the contact point. As seen in Fig. 9A, this corresponds to a critical toppling angle defined by:

$$tan\theta_c = \frac{R}{L-r_{CM}} \tag{13a}$$

With $r_{CM} = r_{CM}(t=0)$ given by Eq. (3e), we obtain:

$$\theta_c = atan\frac{R/L}{(1-\epsilon)\frac{h}{2L}+\frac{\epsilon}{2}} \tag{13b}$$

Fig. 9b shows that the variation of $\theta_c$ as a function of $\widetilde{M}$ is non monotonic when the fluid is at rest. Naively, we could expect that the more water, the higher the centre of mass and the lower the value of $\theta_c$. However, for $M_W < M_b$, the center of mass of the 'Bottle+Water' system becomes mostly set by $M_b$, and thus elevated with respect to the base of the bottle, which explains the maximum value $\theta_c$=45° obtained for $\widetilde{M} = 0.14$. Moreover, due to water redistribution during movement, the location of the centre of mass with respect to the bottle base just at the collision with the ground is expected to be higher as compared with the one for the same mass at rest. Indeed, if $r_{CM}(\theta = 2\pi)$ is extracted from the simulated data for each filling fraction, we find that the corresponding dynamic critical toppling angle is lower than the expected static one, except for $\widetilde{M} > 0.75$ where the agreement is good due to negligible water redistribution. Images taken at collision (Fig. 9c) confirm that a significant mass of water is still elevated in the bottle for intermediate $\widetilde{M}$ values at the moment of the collision. At low $\widetilde{M}$ values, water is found at the bottom and along the body of the bottle in its lower half. As a consequence, in this context, water motion during bottle flip tends to elevate the centre of mass of the system and decrease the stability of the bottle upon landing. There is however an additional effect that contributes bottle stabilization. Quantitative explanations are beyond the scope of the present work and would require detailed fluid hydrodynamics considerations. From a qualitative point of view, collision between the falling bottle and the ground can be analysed in the same framework as the famous "stacked ball drop" leading to the amplification of the velocity of the upper ball after the rebound [10], but with different conditions leading to a damping effect. Here, since the collision duration is on the order of few milliseconds, we may neglect the rotation of the bottle and assume a vertical fall. In the case of the drop of two balls on top of each other, there is one collision between the ground and the lower ball, and a second collision between the two balls. Here the fraction of water that is still dispersed in the bottle at the moment of the collision plays the role of the upper ball. In the case of balls, the lower ball transfers its momentum to the small ball that rebounds with a speed that may exceed the free falling speed. Here, we may realistically assume that, in the early stages of collision (before sloshing and possible jet formation [11], i) the collision is completely inelastic, meaning that the two final velocities are equal, on the order of the weak bouncing velocity of the bottle, and ii) that the momentum of the ascending bottle is partly cancelled by the momentum of the descending dispersed fraction of water. Some favourable cases to enhance this damping effect occur when the velocity of the centre of mass of the whole system is negative. Fig. 9c shows that $v_{CM}$ is indeed negative and its magnitude larger than 0.2 m.s$^{-1}$ in the range $\widetilde{M} \approx 0.25 - 0.6$. Remarkably, this range also matches with the conditions for low angular velocity (Fig. 7b).

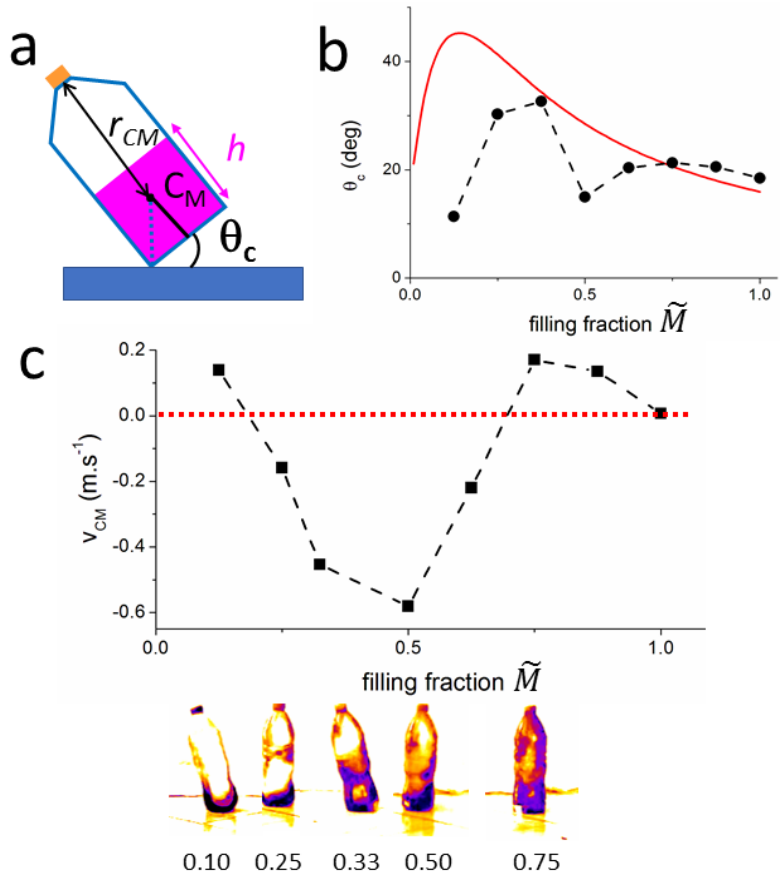

*Figure 9*: *Stabilization effect on landing. (**a**) definition of the critical toppling angle $\theta_c$; (**b**) Variation of $\theta_c$ as a function of filling fraction $\widetilde{M}$. The red curve is computed from Eq. (13b) and obtained from resting value of $r_{CM}$. The data points correspond to the location of the centre of mass on landing; (**c**) Variation of the instantaneous velocity of the centre of mass on landing as a function of filling fraction $\widetilde{M}$ and images of the bottle with fire LUT enhancing the distribution of water inside the bottle.*

## 7. Conclusion

To summarize, we have performed water bottle flipping experiments, and we proposed a simple 1D numerical approach to model the physics of water flipping. The model is based on classical rotational mechanics in a non-inertial frame, which allows undergraduate sutdents to explore with profound understanding the concepts of center of mass, moment of inertia (variable), conservation of angular momentum. The experiments were systematically performed and analyzed in depth. This work is also an example of combination between experimental and theoretical physics, and how both experiments and numerical model can feed each other. In particular, we extensively used the experimental findings to tune the only two phenomenological parameters of the model, which could otherwise look oversimplified. It is quite remarkable that the behavior of a bottle filled with fluid can be well recapitulated by decomposing the water mass into thin slices of rigid body. Finally, beyond the main effect of moment of inertia increas and, concomitantly, angular velocity reduction due to water redistribution, we have also discussed qualitatively the damping effect that stabilizes landing.

**Acknowledgements**

We are grateful to Lucie Nassoy for contributing to the water bottle tosses and recording the experiments performed by the other authors.

# Supplementary material

## A simple numerical model of the water bottle flipping experiment

Julie Nassoy1, Margot Nguyen Huu 1 Léon Rembotte2,3,4,5, Jean-Baptiste Trebbia2,3,4, and Pierre Nassoy2

1 Lycée Montaigne, Bordeaux, France
2 University of Bordeaux, Laboratoire Photonique Numérique et Nanosciences (LP2N), 33400 Talence, France
3 Institut d'Optique, LP2N, 33400 Talence, France
4 CNRS, UMR 5298, 33400 Talence, France
5 Centre de Recherche Paul Pascal (CRPP), UMR 5031, CNRS and University of Bordeaux, 33600 Pessac, France

## Supplementary Figure 1 :

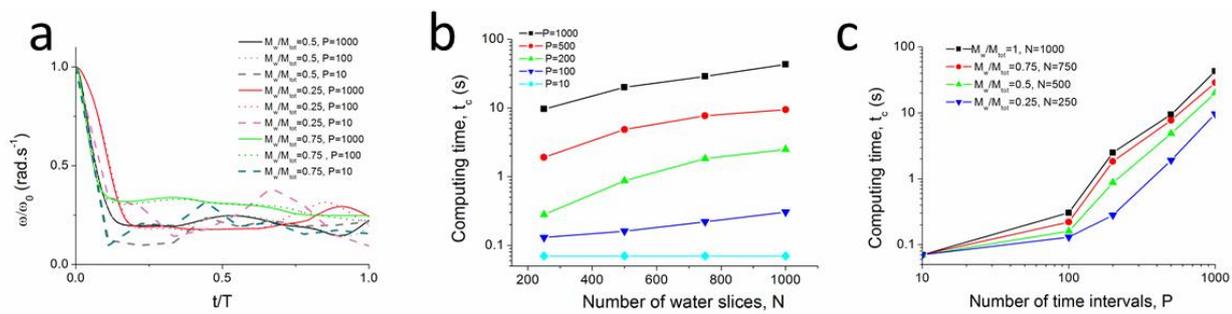

**Supp. Fig. 1** : **(a)** Variation of $\omega/\omega_0$ as a function of t/T for different filling fractions ($\widetilde{M}$=0.25, 0.50 and 0.75) and different numbers of time steps (P=10, 100, and 1000).**(b)** Computing time $t_c$ versus number of slices N (for different P values). **(c)** Computing time $t_c$ versus number of time steps P (for different N values).

## Supplementary Video1:

Water bottle flipping sequence with a filling fraction of 10%. Video is slowed down 9.6 times (from initial frame rate of 240 fps to 25 fps)

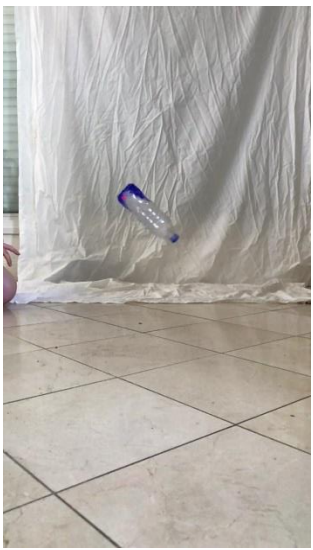

**Supplementary Video2:**

Water bottle flipping sequence with a filling fraction of 25%. Video is slowed down 9.6 times (from initial frame rate of 240 fps to 25 fps)

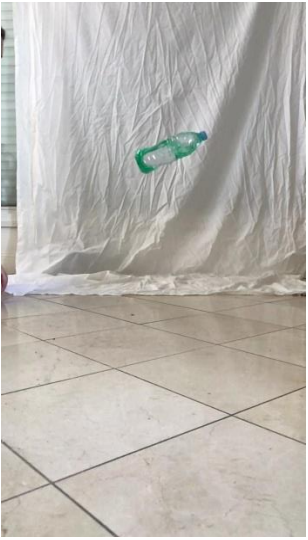

**Supplementary Video3:**

Water bottle flipping sequence with a filling fraction of 33%. Video is slowed down 9.6 times (from initial frame rate of 240 fps to 25 fps)

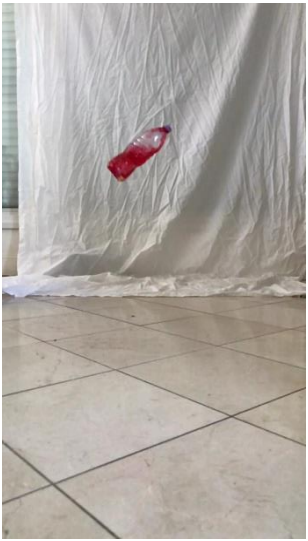

**Supplementary Video4:**

Water bottle flipping sequence with a filling fraction of 50%. Video is slowed down 9.6 times (from initial frame rate of 240 fps to 25 fps)

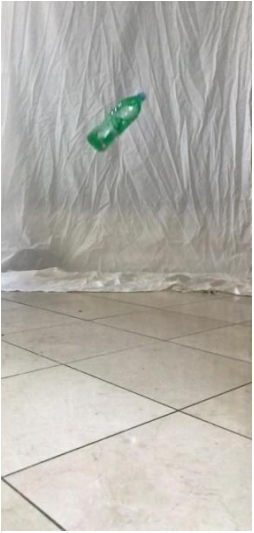

**Supplementary Video5:**

Water bottle flipping sequence with a filling fraction of 75%. Video is slowed down 9.6 times (from initial frame rate of 240 fps to 25 fps)

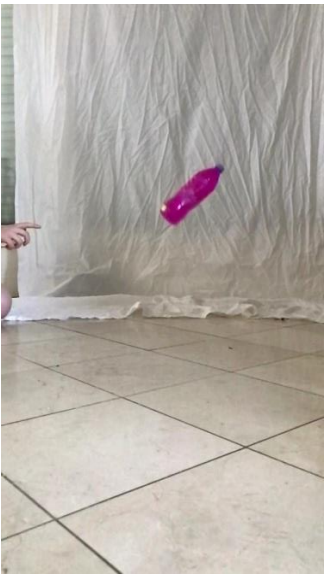

**Supplementary Video6:**

Flipping sequence with a bottle partially filled with ice. Video is slowed down 9.6 times (from initial frame rate of 240 fps to 25 fps)

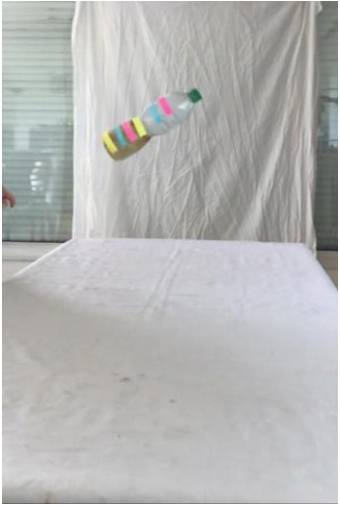

**Supplementary Text:**

*MATLAB program*

**Main.m**

```
% Compute the trajectories of water slices, the center of mass,
% the angle of the bottle during flipping and the angular velocity

%initialisation
tic,
theta =pi/2;  %angle between the bottle and the vertical at t=0
Nmax= 1000; %number of time steps
T=1; %total duration of the flipping motion
t=linspace(0,T,Nmax);
dt =t(2)-t(1);
N=500; %number of water slices
L=0.28; %height of the bottle
R=0.04; %radius of the bottle
g=9.8; %gravity
M_b=0.028; %mass of the bottle
M_max=1; %maximal water in the bottle
M_W =0.50; % mass of water in the bottle
Mtilde=M_W/M_max; %filling fraction
h=0.14; %height of the water level from the bottom of the bottle
rho =1000; density of water
H0 = M_b/(rho*pi*R^2);
epsilon = H0/(H0+h); %correction to take into account the mass of the bottle
alpha=0.01; %empirical drag coefficient
omega = 20; %angular velocity
omega_0 = omega; %angular velocity at time t=0

%vector of the slice position
r = L-(2*(1:N).'-1)*h/(2*N);
%position of the center of mass
r_CM=[];
%compute  the center of mass at time t=0
r_CM0 = FindCenterOfMass(r_CM,r,L,epsilon,1);
%compute the linear velocity at time t=0 from angular velocity
v= 0*r_CM0*omega_0*(1:N).';
%compute the moment of inertia of the water inside the bottle
% with respect to the center of mass of the bottle+water system
JW_CM0= InertiaMomentumWater(r,r_CM0,M_W,N,1,Mtilde,L,R,epsilon);
%compute the moment of inertia of the empty bottle with respect
% to the center of mass of the bottle+water system
JB_CM0 = InertiaMomentumBottle(r_CM0,M_b,R,L,1);
%compute the total moment of inertia of the bottle+water system
J_0 = JW_CM0+JB_CM0;
```

```matlab
% Make a loop on time steps
for n=1:length(t)
    
    %compute the center of mass
    r_CM = FindCenterOfMass(r_CM,r,L,epsilon,n);
    %compute the moment of inertia of the water
    JW_CM= InertiaMomentumWater(r,r_CM,M_W,N,n,Mtilde,R,L,epsilon);
    %compute the moment of inertia of the bottle
    JB_CM= InertiaMomentumBottle(r_CM,M_b,R,L,n);
    %compute the moment of inertia of the bottle+water system
    J=JW_CM+JB_CM;
    %compute the angular velocity
    omega = InstantaneousAngularFrequency(J_0,omega,omega_0,J,n);
    %compute the angle
    theta=FindTheta(theta,omega,dt,n);
    %compute the  velocity and position of slices
    [r,v]=FindNewlocation(r,v,r_CM,dt,g,omega,theta,n,alpha,M_W,N);
    %check the boundary condition
    [r,v]=CheckBoundaryConditions(r,v,h,N,L);
    
end
toc
close all

% Plot figures

% angle versus time
figure
plot(t,theta(1:end-1),'o-')
%center of mass versus time
figure
plot(t,r_CM(1:end))
% angular velocity versus time
figure
plot(t,omega)
% Position of all slices versus time
figure
plot(t,r(1,1:end-1))
hold on
plot(t,r(end,1:end-1))
plot(t,r(round(N/2),1:end-1))
% Position of 1st, (N/2)th and Nth slice versus time
figure
plot(t,r(:,1:end-1))

%Create data file (t, position center of mass, theta, omega)
temporary = [t.'  r_CM(1:end)' theta(1:end-1).' omega.'];
save('C:\Users\bottleflip\file1.txt','temporary','-ascii')
```

**FindCenterOfMass.m**

```
function r_CM = FindCenterOfMass(r_CM,r,L,epsilon,n)
% find the center of mass from the position of the N slices and the
prarmeters L and epsilon
N=size(r,1);
r_CM(n) = (1-epsilon)*sum(1/N*r(:,n),1)+epsilon*L/2;
```

**InertiaMomentumWater.m**

```
function JW_CM= InertiaMomentumWater(r,r_CM,M_W,N,n,Mtilde,R,L,epsilon)
% compute the moment of inertia of the water with respect to the center of
mass of the bottle+water system
%r : position of the N slices
%r_CM: position of the center of mass
JW_CM = M_W/N*sum((r(:,n)-r_CM(n)).^2,1);
```

**InertiaMomentumBottle.m**

```
function JB_CM = InertiaMomentumBottle(r_CM,M_b,R,L,n)
% compute the moment of inertia of the bottle with respect to the center of
mass
JB_CM(n) = M_b*(R^2/2+L^2/12+(L/2-r_CM(n)).^2);
```

**InstantaneousAngularFrequency.m**

```
function omega = InstantaneousAngularFrequency(J_0,omega,omega_0,J,n)
%compute the Instantaneous Angular velocity from the conservation of the
angular momentum
omega(n) = J_0*omega_0/J(n);
```

**FindTheta.m**

```
function theta=FindTheta(theta,omega,dt,n)
%find the new theta after dt at step n
theta(n+1)=theta(n)+dt*omega(n);
```

**FindNewlocation.m**

```
function [r,v] = FindNewlocation(r,v,r_CM,dt,g,omega,theta,n,alpha,M_W,N)
%compute the new locations and speeds of the N  slices at the iteration n
% r and v are the position and velocity
%r_CM is the center of mass
%omega:  angular velocity
%theta: angle of the bottle
%g gravity
%dt: step time

r(:,n+1) = (r(:,n)-r_CM(n)+g/omega(n)^2.*cos(theta(n)))*exp(-
alpha/2/M_W*N*dt)*cosh(omega(n)*dt)+(v(:,n)/omega(n)+alpha/2/M_W/N/omega(n)*
(r(:,n)-r_CM(n)+g/omega(n)^2.*cos(theta(n))))*exp(-
alpha/2/M_W*N*dt)*sinh(omega(n)*dt)...
```

```
    +r_CM(n)-g/omega(n)^2*cos(theta(n));

v(:,n+1) = ((r(:,n)-r_CM(n)+g/omega(n)^2.*cos(theta(n)))*(omega(n)^2-
(alpha/2/M_W*N)^2)/omega(n)+alpha/2/M_W*N/omega(n)*v(:,n))*exp(-
alpha/2/M_W*N*dt)*sinh(omega(n)*dt)+v(:,n)*cosh(omega(n)*dt)*exp(-
alpha/2/M_W*N*dt);
```

**CheckBoundaryConditions.m**

```
function [r_new,v_new]=CheckBoundaryConditions(r,v,h,N,L)

N=size(r,1);
%slices that do not hit the bottom of top of the bottle are unchanged
current_v = v(:,end);
current_r = r(:,end);
% If slice touches the ends of the bottle, they bounce back with a new
velocity equal to -1/2 of the previous velocity (value beta in the text)
current_v(current_r<(h/(2*N))) = -0.5*current_v(current_r<(h/(2*N)));
current_v(current_r>(L-h/(2*N))) = -0.5*current_v(current_r>(L-h/(2*N)));
v_new = [v(:,1:end-1) current_v];
r_new = r;
```

---

*PYTHON program*

**main.py**

```python
# Import modules
import numpy as np
import matplotlib.pyplot as plt
import kinetics
from constants import *

# Set kinetic parameters
water_mass = 0.1  # Mass of water in the bottle, unit: kg
filling_fraction = water_mass / WATER_MASS_MAX  # Filling fraction of the
bottle
water_height = filling_fraction * BOTTLE_HEIGHT  # Height of the water level
from the bottom of the bottle at rest, unit: m
epsilon = BOTTLE_MASS / (BOTTLE_MASS + water_mass)  # correction factor
alpha = 0.01  # Empirical drag coefficient, unit: kg/m
beta = -0.3  # Empirical restitution coefficient, unit: 1
omega_0 = 20  # Initial angular velocity, unit: rad/s
theta_0 = np.pi / 2  # Initial angle, unit: rad

# Set numerical parameters for the simulation
n_steps = 1000  # Number of time steps
```

```python
t_max = 0.6  # Total duration of the flipping motion, unit: s
t, dt = np.linspace(0, t_max, n_steps, retstep=True)  # Time vector and time step
n_slices = 100  # Number of water slices
slice_height = water_height / n_slices  # Height of a slice, unit: m
slice_mass = water_mass / n_slices  # Mass of a slice, unit: kg

# Initialize the variables
omega = omega_0  # Initial angular velocity, unit: rad/s
theta = theta_0
slices_positions = BOTTLE_HEIGHT - slice_height / 2 - np.arange(0, n_slices) * slice_height
slices_velocities = 0 * slices_positions  # Initial radial velocity of the slices
center_of_mass_system = kinetics.find_center_of_mass(slices_positions, epsilon)
rotational_inertia_water = kinetics.rotational_inertia_water(slices_positions, center_of_mass_system, water_mass, n_slices)
rotational_inertia_bottle = kinetics.rotational_inertia_bottle(center_of_mass_system, BOTTLE_MASS, BOTTLE_RADIUS, BOTTLE_HEIGHT)
rotational_inertia = rotational_inertia_water + rotational_inertia_bottle

# Create vectors to store the results
omega_values = np.zeros(n_steps)
theta_values = np.zeros(n_steps)
slices_positions_values = np.zeros((n_steps, n_slices))
slices_velocities_values = np.zeros((n_steps, n_slices))
center_of_mass_system_values = np.zeros(n_steps)
rotational_inertia_values = np.zeros(n_steps)

# Store initial conditions
omega_values[0] = omega
theta_values[0] = theta
slices_positions_values[0] = slices_positions
slices_velocities_values[0] = slices_velocities
center_of_mass_system_values[0] = center_of_mass_system
rotational_inertia_values[0] = rotational_inertia

# Numerical integration
for k in range(1, len(t)):

    # Recall values of the variables
    omega = omega_values[k - 1]
    theta = theta_values[k - 1]
    slices_positions = slices_positions_values[k - 1]
    slices_velocities = slices_velocities_values[k - 1]
    center_of_mass_system = center_of_mass_system_values[k - 1]
```

```python
        rotational_inertia = rotational_inertia_values[k - 1]

        # Update values of the variables using the previous values
        slices_positions_tmp, slices_velocities_tmp = 
kinetics.update_slice_positions(slices_positions, center_of_mass_system, 
omega, theta, slices_velocities, alpha, slice_mass, dt, G)
        new_slices_positions_tmp, new_slices_velocities_tmp = 
kinetics.check_boundary_conditions(slices_positions_tmp, slices_positions, 
slices_velocities_tmp, beta, slice_height / 2, BOTTLE_HEIGHT - slice_height 
/ 2)
        center_of_mass_system_tmp = 
kinetics.find_center_of_mass(slices_positions, epsilon)
        rotational_inertia_water_tmp = 
kinetics.rotational_inertia_water(slices_positions, center_of_mass_system, 
water_mass,n_slices)
        rotational_inertia_bottle_tmp = 
kinetics.rotational_inertia_bottle(center_of_mass_system, BOTTLE_MASS, 
BOTTLE_RADIUS, BOTTLE_HEIGHT)
        rotational_inertia_tmp = rotational_inertia_water_tmp + 
rotational_inertia_bottle_tmp
        omega_tmp = omega_values[0] * rotational_inertia_values[0] / 
rotational_inertia_tmp
        theta_tmp = theta + omega * dt

        # Store new values of the variables
        omega_values[k] = omega_tmp
        theta_values[k] = theta_tmp
        slices_positions_values[k] = new_slices_positions_tmp
        #print(slices_positions_values)
        slices_velocities_values[k] = new_slices_velocities_tmp
        center_of_mass_system_values[k] = center_of_mass_system_tmp
        rotational_inertia_values[k] = rotational_inertia_tmp

# Plot figures
fig, ax = plt.subplots(3, 1, figsize=(5, 10))
ax[0].plot(t, theta_values)
ax[0].set_xlabel('Time')
ax[0].set_ylabel('Angle')

ax[1].plot(t, center_of_mass_system_values, '--')
ax[1].plot(t, slices_positions_values[:, 0], 'k')
ax[1].plot(t, slices_positions_values[:, -1], 'k')
ax[1].plot(t, slices_positions_values[:, n_slices // 2], 'k')
ax[1].set_xlabel('Time')
ax[1].set_ylabel('Center of Mass')
ax[1].set_ylim(0, 1.1 * BOTTLE_HEIGHT)

ax[2].plot(t, omega_values)
ax[2].set_xlabel('Time')
```

```python
ax[2].set_ylabel('Angular Velocity')
ax[2].set_ylim(0, 1.1 * omega_0)
plt.tight_layout()
plt.show()

# Create data file (t, position center of mass, theta, omega)
# data = np.array([t, center_of_mass_system_values, theta_values,
omega_values]).T
# np.savetxt('data.txt', temporary_data, delimiter=' ')
```

**Constants.py**

```python
# Declare constants
BOTTLE_HEIGHT = 0.28  # Height of the bottle, unit: m
BOTTLE_RADIUS = 0.04  # Radius of the bottle, unit: m
BOTTLE_MASS = 0.028  # Mass of the bottle, unit: kg
WATER_MASS_MAX = 1  # Maximal water in the bottle, unit: kg
G = 9.81  # Gravity constant on Earth, unit: m/s**2
# WATER_DENSITY = 1000  # Density of water, unit: kg/m**3
```

**Kinetics.py**

```python
# Import modules
import numpy as np
from constants import *

# Compute the rotational inertia of the water with respect to the center of
mass
def rotational_inertia_water(positions, center_of_mass,
water_mass,n_slices):
    #j = np.sum(mass * (positions - center_of_mass) ** 2)
    j = water_mass/n_slices*np.sum((positions - center_of_mass) ** 2)
    return j

# Compute the rotational inertia of the bottle with respect to the center of
mass
def rotational_inertia_bottle(center_of_mass, mass_bottle=BOTTLE_MASS,
radius=BOTTLE_RADIUS, length=BOTTLE_HEIGHT):
    j_bottle_axis = mass_bottle * (radius ** 2 / 2 + length ** 2 / 12)
    j_new_axis = j_bottle_axis + mass_bottle * (length / 2 - center_of_mass)
** 2  # Parallel axis theorem
    return j_new_axis

# Compute the position of the center of mass
def find_center_of_mass(positions, eps, length_bottle=BOTTLE_HEIGHT):
    r_center_of_mass = eps * length_bottle / 2 + (1 - eps) *
np.mean(positions)
    return r_center_of_mass
```

```python
# Update slice positions and velocities using previous values
def update_slice_positions(positions, center_of_mass, angular_velocity,
angle, velocities, alpha, mass, dt, gravity):
    tau = (2 * mass) / alpha
    factor_1 = positions - center_of_mass + (gravity / (angular_velocity **
2)) * np.cos(angle)
    factor_2 = velocities / angular_velocity + (1 / (tau *
angular_velocity)) * (positions - center_of_mass + (gravity /
(angular_velocity ** 2)) * np.cos(angle))
    new_positions = center_of_mass - (gravity / (angular_velocity ** 2)) *
np.cos(angle) + factor_1 * np.exp(- dt / tau) * np.cosh(angular_velocity *
dt) + factor_2 * np.exp(- dt / tau) * np.sinh(angular_velocity * dt)
    factor_1 = ((angular_velocity ** 2 - (1 / (tau ** 2))) /
angular_velocity) * (positions - center_of_mass + (gravity /
(angular_velocity ** 2)) * np.cos(angle)) + (1 / (tau * angular_velocity))
*velocities
    factor_2 =  velocities
    new_velocities = factor_1* np.exp(- dt / tau) * np.sinh(angular_velocity
* dt) + factor_2 * np.exp(- dt / tau) * np.cosh(angular_velocity * dt)
    return new_positions, new_velocities

# Take into account the rebound on the boundaries
def check_boundary_conditions(positions, previous_positions, velocities,
restitution, l_min, l_max):
    for i in range(len(positions)):
        if positions[i] < l_min:
            velocities[i] = restitution * velocities[i]
            positions[i] = previous_positions[i].copy()  # Just use the
previous position
            #positions[i] = l_min + (l_min - positions[i])  # Would make
more sense to have a symmetrical rebound
        elif positions[i] > l_max:
            velocities[i] = restitution * velocities[i]
            positions[i] = previous_positions[i].copy()  # Just use the
previous position
            #positions[i] = l_max - (positions[i] - l_max)  # Would make
more sense to have a symmetrical rebound
    return positions, velocities
```